\documentclass[a4paper]{jpconf}
\usepackage{graphicx}
\usepackage{amssymb}

\newcommand{\OmM}{\Omega_{\rm M}}
\newcommand{\OmL}{\Omega_\Lambda}
\newcommand{\OmB}{\Omega_{\rm B}}
\newcommand{\Ho}{H_0}
\newcommand{\CTT}{C^{\rm TT}}
\newcommand{\CTE}{C^{\rm TE}}
\newcommand{\CEE}{C^{\rm EE}}
\newcommand{\CBB}{C^{\rm BB}}
\newcommand{\EI}{E_{\rm I}}
\newcommand{\Mpl}{M_{\rm pl}}

\begin{document}
\title{Cosmic Microwave Background Polarization}

\author{James G. Bartlett}

\address{APC, 11 pl. Marcelin Berthelot, 75231 
             Paris Cedex 05, FRANCE\\	
		 (UMR 7164 CNRS, Universit\'e Paris 7, CEA, Observatoire de Paris)}

\ead{bartlett@apc.univ-paris7.fr}

\begin{abstract}
Cosmic microwave background (CMB) anisotropy is our richest
source of cosmological information; the standard cosmological
model was largely established thanks to study of the temperature
anisotropies.  By the end of the decade, the Planck satellite will
close this important chapter and move us deeper into the new
frontier of polarization measurements.  Numerous ground--based
and balloon--borne experiments are already forging into this
new territory.  Besides providing
new and independent information on the primordial density 
perturbations and cosmological parameters, polarization measurements
offer the potential to detect primordial gravity waves, constrain
dark energy and measure the neutrino mass scale.  A vigorous 
experimental program is underway worldwide and heading towards 
a new satellite mission dedicated to CMB polarization.
\end{abstract}

\section{Introduction}

Observations of the cosmic microwave background (CMB) anisotropy have 
driven the remarkable advance of cosmology over the past 
decade~\cite{cmb_t}.  
They tell us that we live in a spatially flat universe where
structures form by the gravitational evolution of nearly scale 
invariant, adiabatic perturbations in a predominant 
form of non--baryonic cold dark matter. Together with either 
results from supernovae Ia (SNIa) distance measurements~\cite{sn},
the determination of the Hubble constant~\cite{freedman01}
or measures of large scale structure~\cite{lss}, they furthermore 
demonstrate that a mysterious dark energy (cosmological constant, 
vacuum energy, quintessence...) dominates the total energy density of our Universe.
These observations have established what is routinely called 
the standard cosmological model: $\OmM\approx 0.3=1-\OmL$, 
$\OmB h^2\approx 0.024$ and $\Ho\approx 70$~km/s/Mpc~\cite{stand_model}.  
Because the observations in fact over--constrain the model, they
test its coherence and its foundations, marking a new era in cosmology.

The CMB results are remarkable for several reasons. They show
us density perturbations on superhorizon scales at decoupling and 
therefore evidence for new physics (inflation or other) working in the
early universe.  The observed peaks in the power spectrum confirm the 
key idea that coherent density perturbations enter the horizon
and begin to oscillate as acoustic waves in the primordial plasma
prior to recombination; their position justifies the
long--standing theoretical preference for flat space with zero curvature.
Their heights measure both the total matter 
and baryonic matter densities, and thereby provide direct evidence that most
of the matter is non--baryonic; and, in a scientific {\em tour de force},
the CMB--determined baryon density broadly agrees 
with the totally independent estimation from Big Bang 
Nucleosynthesis~\cite{bbn}.

These milestones were all obtained from study of the temperature,
or total intensity, anisotropies.  The Planck 
mission\footnote{\tt http://www.esa.int/science/planck} (launch 2007/2008)
will largely complete this work by decade's end with foreground--limited 
temperature maps down to $\sim 5$~arcmin resolution, leaving only the 
smallest scales unexplored.  With this in mind, the field is already
turning to CMB polarization measurements and their wealth of new 
information.

\section{Polarization}

Thomson scattering generates CMB polarization anisotropy at 
decoupling~\cite{cmb_polar}.
This arises from the polarization dependence of the differential
cross section: $d\sigma/d\Omega\propto |\epsilon'\cdot\epsilon|^2$,
where $\epsilon$ and $\epsilon'$ are the incoming and outgoing 
polarization states~\cite{rad_basics}; only linear polarization is involved.  
This dependence means that an observer measuring a given polarization 
sees light scattered preferentially from certain directions around the
scattering electron in the last scatter surface.  The 
orthogonal polarization preferentially samples different parts
of the sky.  Any local intensity anisotropy around the scattering
electron thus creates a net linear polarization at the observer's 
detector.  Quantitatively, only a local quadrupolar temperature
anisotropy produces a net polarization, because of the $\cos^2\theta$
dependence of the cross section.  Also notice that the signal 
is generated in the last scattering surface, where the optical
depth transits from large to small values; the optical depth
must of course be non--zero, but too large a value would 
erase any local anisotropy.  

\begin{figure}[h]
\begin{minipage}{18pc}
\includegraphics[width=18pc]{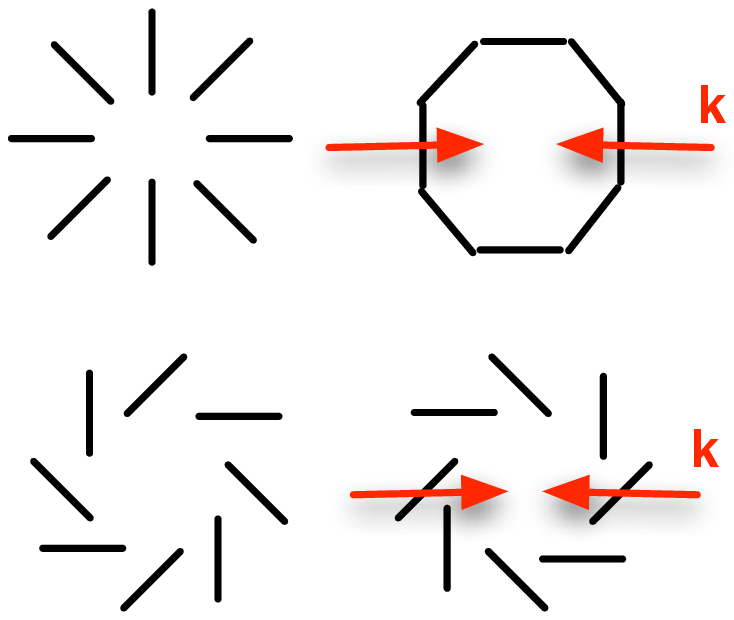}
\caption{\label{fig:eb_patterns}Polarization patterns around
an local intensity extremum.  The upper row shows the even parity
$E$--mode (negative on the left, positive on the right), and the
lower row the odd parity $B$--mode (negative on the left, positive
on the right).  The red arrows illustrate the projected wave vectors
of plane wave perturbations converging at the extremum.}
\end{minipage}\hspace{2pc}%
\begin{minipage}{18pc}
\includegraphics[width=18pc]{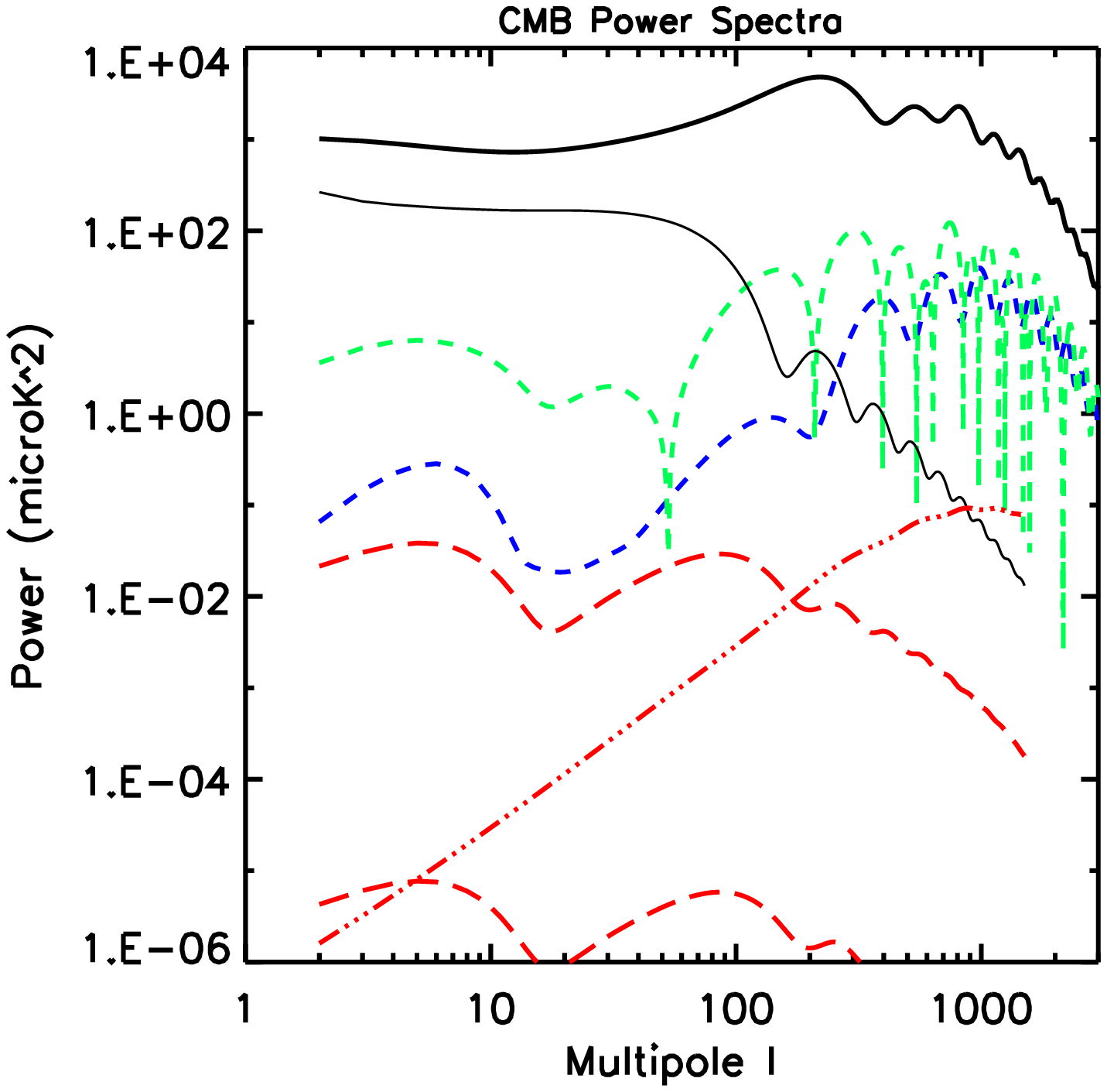}
\caption{\label{fig:spectra}Angular power spectra 
\mbox{$l(l+1)C_l/2\pi$}.  The bold solid black line
shows the temperature power spectrum of the standard model~\cite{stand_model}; the thin black line gives the tensor
contribution to the temperature power for $r=0.5$.  The green (upper) and 
blue (lower) short dashed curves are, respectively, the scalar $TE$ (absolute 
value shown) and $EE$ power 
spectra for the standard model; the former is well measured on large
scales by WMAP~\cite{wmap_polar}.  The red long dashed lines indicate the tensor $B$--mode power for $r=0.5$ (upper) and $r=10^{-4}$ (lower).  Gravitational lensing produces the $B$--mode power shown as the red 
3--dot--dashed curve peaking at $l\sim 1000$.}
\end{minipage} 
\end{figure}

\subsection{Describing CMB Polarization}
Polarized light is commonly described with the Stokes 
parameters~\cite{rad_basics}.
As we have seen, the CMB is linearly polarized, so we use
the Stokes parameters $Q$ and $U$, each of which is defined as the 
intensity difference between two orthogonal polarization directions.
Let $(x,y)$ and $(x',y')$ refer to two coordinate systems situated perpendicular to the light propagation direction and rotated by
$45$ degrees with respect to each other.  Then $Q\equiv I_y-I_x$
and $U\equiv I_{y'}-I_{x'}$.  

Clearly, the values of $Q$ and $U$ will depend on the orientation 
of the coordinate system used at each point on the sky.  Although
from an experimental viewpoint this is unavoidable, it is better
for theoretical purposes to look for a coordinate--free
description; this latter could then be translated into any chosen
coordinate system.  Two such descriptions were first proposed for
the CMB by Zaldariagga \& Seljak and by Kamionkowski et 
al.~\cite{polar_decomp}.  The former, in particular,
model polarization as a spin 2 field
on the sphere, an approach used in the publically
available CMB codes~\cite{cmb_codes}.  

The coordinate--free description distinguishes two kinds of 
polarization pattern on the sky by their different parities. 
In the spinor approach, the even parity 
pattern is called the $E$--mode and the odd parity pattern 
the $B$--mode.  We can represent the polarized CMB sky by 
a map color--coded for intensity and with small bars indicating 
the direction of linear polarization at each point.  Consider
a peak in the intensity (see Figure~\ref{fig:eb_patterns}).  
If the polarization bars are oriented either
in a tangential or a radial pattern around the peak, we have a
$E$--mode; if they are oriented at $45$~degrees (relative to rays 
emanating from the peak), we have $B$--mode:  a reflection of the
sky about any line through the peak leaves the $E$--mode unchanged
(even parity), while the $B$--mode changes sign (odd parity)\footnote{These
local considerations generalize to the sphere~\cite{polar_decomp}.}. 

Another useful way to see this is to consider the wave vectors  
of the plane wave perturbations making up the intensity peak; they
radially point towards the peak center (see Figure).  We then see that
an $E$--mode plane wave has its polarization either perpendicular
or parallel to the wave vector.  A $B$--mode plane wave, on the other
hand, has a linear polarization at $45$~degrees to the wave vector.
The wave vector in fact defines a natural coordinate system
for definition of the Stokes parameters: in this system, $Q$=$E$ and
$U$=$B$.  This is particularly useful when discussing interferometric
observations. 

\subsubsection{The $E$--$B$ Decomposition}
This decomposition of polarization into $E$ and $B$--modes is 
powerful and practical.  As we shall see below, the two different modes 
are generated by different physical mechanisms, which is not surprising,
since they are distinguished by their parity.  Secondly, their different
parity also guaranties that we can separate and individually measure 
the two modes and the total intensity on the sky.  This is extremely 
important because the intensity and polarization anisotropies have very 
different amplitudes (see Figure~\ref{fig:spectra}).  

Theory predicts that the primary CMB anisotropy is a Gaussian field 
(of zero mean), and current observations remain fully consistent with 
this expectation.  We therefore describe CMB anisotropy with the 
power spectrum $C_l$, which is nothing other than the second moment
of the field in harmonic space (i.e., the variance).  As stated,
most of the CMB milestones have been obtained from temperature 
measurements, which in this context means from measurement of the
temperature angular power spectrum $\CTT_l$.  With the introduction
of polarization, we see that in fact there are a total of 4 
power spectra to determine: $\CTT, \CTE, \CEE, \CBB$; parity
considerations eliminate the two other possible power 
spectra\footnote{The cosmological principle prohibits
any preferred parity in the clustering hierarchy, implying that statistical 
measures of the primordial anisotropy have even parity.}, 
$C^{TB}$ and $C^{EB}$.

\subsection{The Physical Content of CMB Polarization}
In the standard model, inflation generates both scalar (S), or density
perturbations~\cite{dens_perts_inflation} and tensor (T), or gravity wave perturbations~\cite{gw_inflation}. 
The scalar perturbations are created by quantum fluctuations in the
particle field (usually assumed to be a scalar field) driving inflation.
After inflation, these perturbations grow by gravity to form galaxies 
and the observed large scale structure.  Gravity waves, on the other hand,
decay once they enter the horizon, and thus leave their imprint in the 
CMB on large angular scales (around and larger than the decoupling
horizon, $\sim 1\deg$).

Gravity wave production by inflation, although a reasonable extrapolation of 
known physics, would nevertheless be something fundamentally new.
These waves would not be generated by any classical or even quantum
source (i.e., by the right--hand--side of Einstein's equations); 
we suppose instead that the gravitational field itself (more specifically,
the two independent polarization states of a free gravity wave in flat space)
experiences vacuum quantum fluctuations like a scalar field.  Finding  
gravity waves from inflation would therefore not simply be a detection
of gravity waves, but also a remarkable observation of the 
semi--classical behavior of gravity.

Both scalar and tensor perturbations contribute to the temperature 
power spectrum; in practice, however, the scalar mode dominates, so that
the measured temperature spectrum effectively fixes the scalar perturbation amplitude.  We quantify the relative amplitude of the scalar and tensor perturbations by the parameter 
$r\equiv {\cal P}_{\rm T}/{\cal P}_{\rm S}$, where ${\cal P}_{\rm T}$ and 
${\cal P}_{\rm S}$ represent the power in the respective modes at a pivot
wavenumber\footnote{In~\cite{peiris_etal03}, they use
$k=0.002$~Mpc$^{-1}$.}~\cite{r_def_power}.  
Since the scalar power is measured, 
we use $r$ to express the gravity wave amplitude.   
The WMAP data combined with large scale 
structure data limit $r< 0.53$ (95\%~\cite{peiris_etal03}).  
Unlike the scalar modes, which depend 
on the slope of the inflation potential, the gravity wave amplitude 
depends only on the energy scale of inflation, $\EI$ 
(specifically, ${\cal P}_{\rm T} \propto (\EI/\Mpl)^4$, where 
$\Mpl$ is the Planck mass).  
Quantitatively\footnote{The numerical relation refers to the 
definition of $r$ used in the above references and corresponds to
the parameter employed in the CAMB code; it furthermore adopts the
$2\sigma$ upper limit on the scalar power amplitude given by 
WMAP.}, we have $\EI = 3.4\times 10^{16}$~GeV~$r^{1/4}$.  Thus, the 
above limit on $r$ corresponds to an upper limit on the inflation scale of 
$E_I<2.9\times 10^{16}$~GeV.    

In addition to these primary CMB polarization signals, gravitational lensing by structures forming along the line--of--sight to the last scattering surface 
sources a particular kind of secondary polarization signal; I discuss this in the following section.

Figure~\ref{fig:spectra} shows inflationary predictions for the various 
CMB power spectra from inflation--generated scalar and tensor 
perturbations\footnote{These calculations where made using CAMB
({\tt http://www.camb.info})}.
The temperature power spectrum for the standard cosmological model
(fit to the WMAP and other higher resolution
experimental data) is given by the bold solid black curve.  This is 
now measured to the cosmic variance limit out to the second peak.
The light, black solid curve gives the maximum allowable 
tensor contribution to the temperature power spectrum, i.e., 
at the current limit of $r<0.53$.  Note that it would be very hard
to significantly improve on this limit with only temperature measurements. 
This is why there is such intense interest in polarization.

\subsubsection{The Importance of the $E$--$B$ Decomposition}
Although both scalar and tensor perturbations generate temperature 
anisotropy ($\CTT$), primordial scalar perturbations only produce $E$--mode polarization, and hence only contribute to $\CEE$ and $\CTE$. {\em They cannot create $B$--mode polarization}.  We see this by considering a plane wave scalar 
perturbation passing over a scattering electron: the local intensity 
quadrupole around the electron must be aligned with the wave vector, which 
implies that the polarization of the scattered light must be 
either perpendicular or parallel to the projected wave vector -- in other words, 
a pure $E$ mode.  The axial symmetry imposed by the scalar nature of the
density perturbations prevents any $B$--mode production.

The short dashed green (upper) and blue (lower) lines 
show $\CTE$ and $\CEE$ power spectra from scalar perturbations for 
the standard model.  These predictions follow directly from the 
measured temperature power spectrum and the assumption -- usually adopted in
the standard model -- that the scalar perturbations are purely adiabatic.
Given the measured temperature spectrum, we could change the predicted
$\CTE$ and $\CEE$ spectra by adding isocurvature perturbations.  Observations
of these polarization modes will therefore constrain the presence of such isocurvature
modes\footnote{These modes are also constrained by large scale structure observations.}.

The TE cross spectrum in fact changes sign, but since I only plot its absolute value, the curve oscillates between the sharp dips corresponding to 
the sign change.  The additional bump at low multipole arises from reinonization,
for which I have taken an optical depth of $\tau=0.17$~\cite{stand_model}.   

In contrast to scalar perturbations, gravity waves (T) 
push and pull matter in directions perpendicular to their propagation, 
aligning the local intensity 
quadrupole in the plane perpendicular to the wave vector.  The loss
of axial symmetry allows both $E$ and $B$--mode production.  Since
the expansion dampens gravity waves on scales smaller than the horizon,
these tensor effects only appear on angular scales larger than 
$\sim 1^\circ$ (the angular size of the decoupling horizon).  
{\em Hence, $B$--mode polarization on large angular scales is the unique
signature of primordial gravity waves} (the so--called smoking 
gun)~\cite{gw_polar}.

As mentioned, the amplitude of the gravity wave signal depends only on
the energy scale of inflation.  A measurement of $B$--mode polarization
on large scales would give us this amplitude, and hence {\em a direct
determination of the energy scale of inflation.}\footnote{More precisely, 
at the end of inflation.}. The red long--dashed curves in 
Figure~\ref{fig:spectra} show the tensor $B$--mode
spectrum for two different amplitudes -- the upper curve for the 
the current limit of $r<0.5$ ($\EI \sim 3\times 10^{16}$~GeV), and 
the lower one for $r=10^{-4}$ ($\EI\sim 3.4\times 10^{15}$~GeV).

Gravitational lensing of CMB anisotropy by structures forming along the 
line--of--sight to decoupling also generates $B$--mode polarization, 
but on smaller scales~\cite{cmb_lensing_b}.  
Lensing deviates the photon trajectories 
(preserving surface brightness) and scrambles (distorts) our view
of the decoupling surface~\cite{cmb_lensing_t}.  
The $E$--$B$ modes are defined as pure
parity patterns on the sky; scrambling any such pattern will 
clearly destroy its pure parity, thereby leaking power into the opposite
parity mode.  If, for example, there were only $E$--mode perturbations 
at decoupling (e.g., gravity waves are negligible), we would still see
some $B$--mode in our sky maps on small angular scales caused by 
gravitational lensing.

The lensing signal has both negative and positive aspects in the
present context.  On the down side, it masks the gravity wave $B$--mode
with a foreground signal with an identical electromagnetic spectrum; thus, we 
cannot remove it using frequency information.  We can, however, extract and 
remove the lensing signal by exploiting the unique mode--mode coupling 
(between different multipoles, absent in the primary anisotropies) induced
by the lensing~\cite{lensing_reconstruction}.  
Uncertainty in this process may ultimately limit our 
sensitivity to gravity waves~\cite{gw_limit_lensing}.

On the positive side, the lensing signal carries important information about
the matter power spectrum and its evolution over a range of redshift 
inaccessible by any other observation.  This provides us with a powerful 
means of constraining dark energy and a singular method for determining  
the neutrino mass scale~\cite{cmb_numass_de}.  
Since the expansion governs the matter perturbation
growth rate, comparison of the amplitude of the power spectrum at high redshift 
to its amplitude today probes the influence of dark energy.  The shape
of the power spectrum, on the other hand, is affected by the presence of
massive neutrinos, which tend to smooth out perturbations by free streaming
out of over-- and under--densities.  The effect 
suppresses the power spectrum on small scales.  

Recent studies indicate that by measuring the lensing polarization signal
to the cosmic variance limit, we would obtain a $1\sigma$ 
sensitivity to the sum of the 3 neutrino masses of 
$\sigma_\Sigma=0.035$~eV~~\cite{cmb_numass_de}.  This is extremely important:
current neutrino oscillation data call for a 
$\Delta m^2 = (2.4_{-0.6}^{+0.5})\times 10^{-3}$~eV$^2$ 
($2\sigma$)~\cite{D_nu2}\footnote{Here, $\Delta m^2$ is the difference between 
the singlet neutrino mass squared and the mean squared mass of the neutrino doublet; see reference.},
implying that the summed mass of the three neutrino species exceeds 
the ultimate CMB sensitivity.
{\em CMB polarization therefore provides a powerful and unique way to measure 
the neutrino mass scale, down to values unattainable in the laboratory.}

The red triple--dot--dashed curve in Figure~\ref{fig:spectra} shows the 
$B$--mode polarization from lensing.  Its amplitude is set by the 
amplitude of the primordial $E$--mode signal and of the matter power 
spectrum as it evolves.  Since gravity waves generate both $E$--modes 
(the tensor contribution is not shown in the figure) and $B$--modes 
of roughly equal power, we expect the scalar $E$--mode to dominate.
Thus, we have a good idea of the overall amplitude of the lensing
$B$--mode spectrum, although the exact amplitude and shape will 
depend, as discussed, on the presence of isocurvature modes, neutrinos and
the nature of dark energy.  For the curve shown in 
Figure~\ref{fig:spectra}, I have adopted the standard model 
(no isocurvature perturbations) with a pure cosmological constant 
and have ignored neutrinos.   

\begin{figure}[h]
\begin{minipage}{18pc}
\includegraphics[width=18pc]{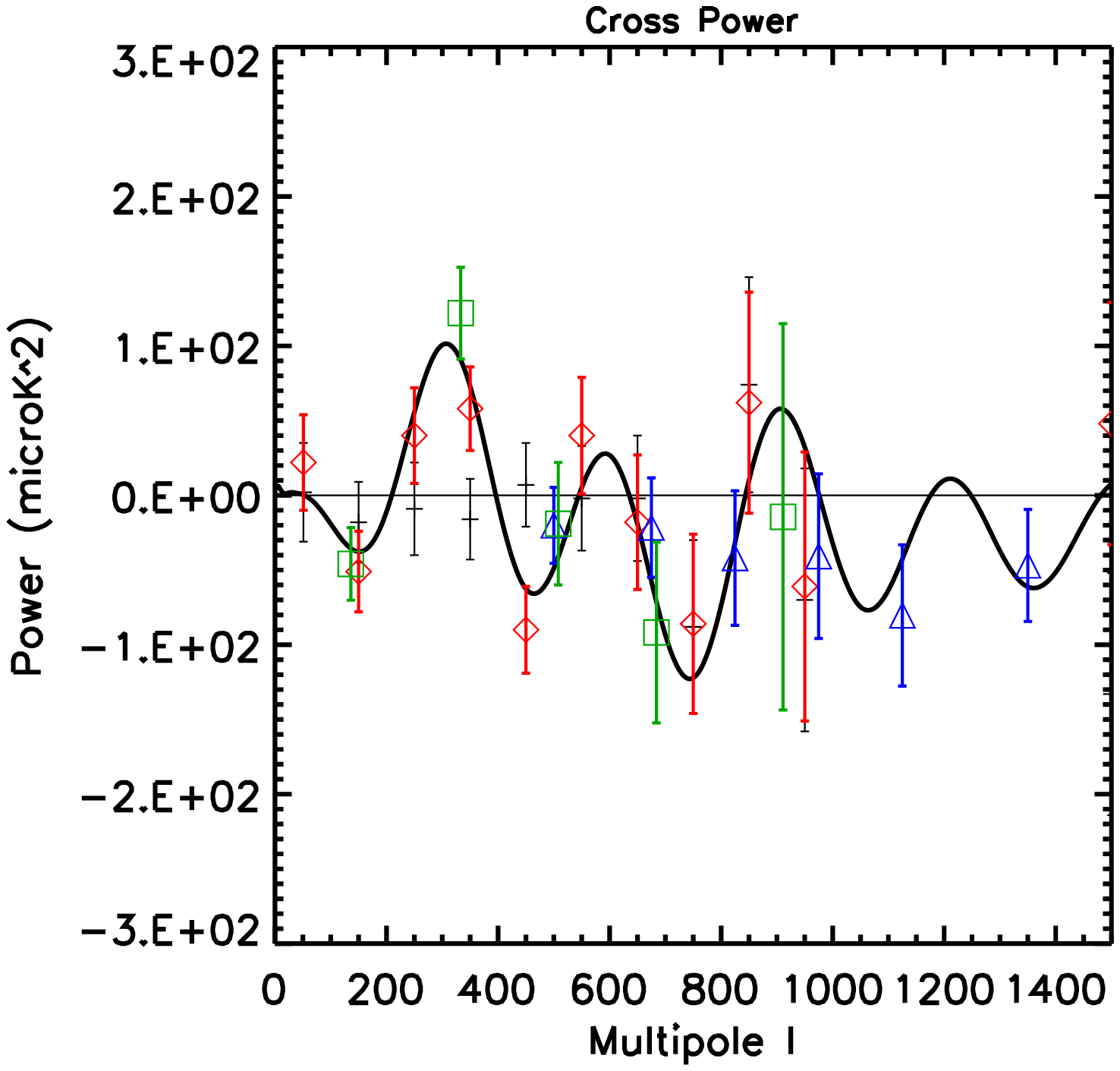}
\caption{\label{fig:te}TE power spectra \mbox{$l(l+1)C_l/2\pi$}.  
The curve shows the power
predicted by the standard model (and measured on large scales by 
WMAP~\cite{wmap_polar}, although not reproduced here).  Red diamonds 
give the BOOMERanG results, green boxes the DASI 3--year results and
blue triangles the CBI results.   
The thin black error bars show
the BOOMERanG $TB$ power, a foreground tracer.}
\end{minipage}\hspace{2pc}%
\begin{minipage}{18pc}
\includegraphics[width=18pc]{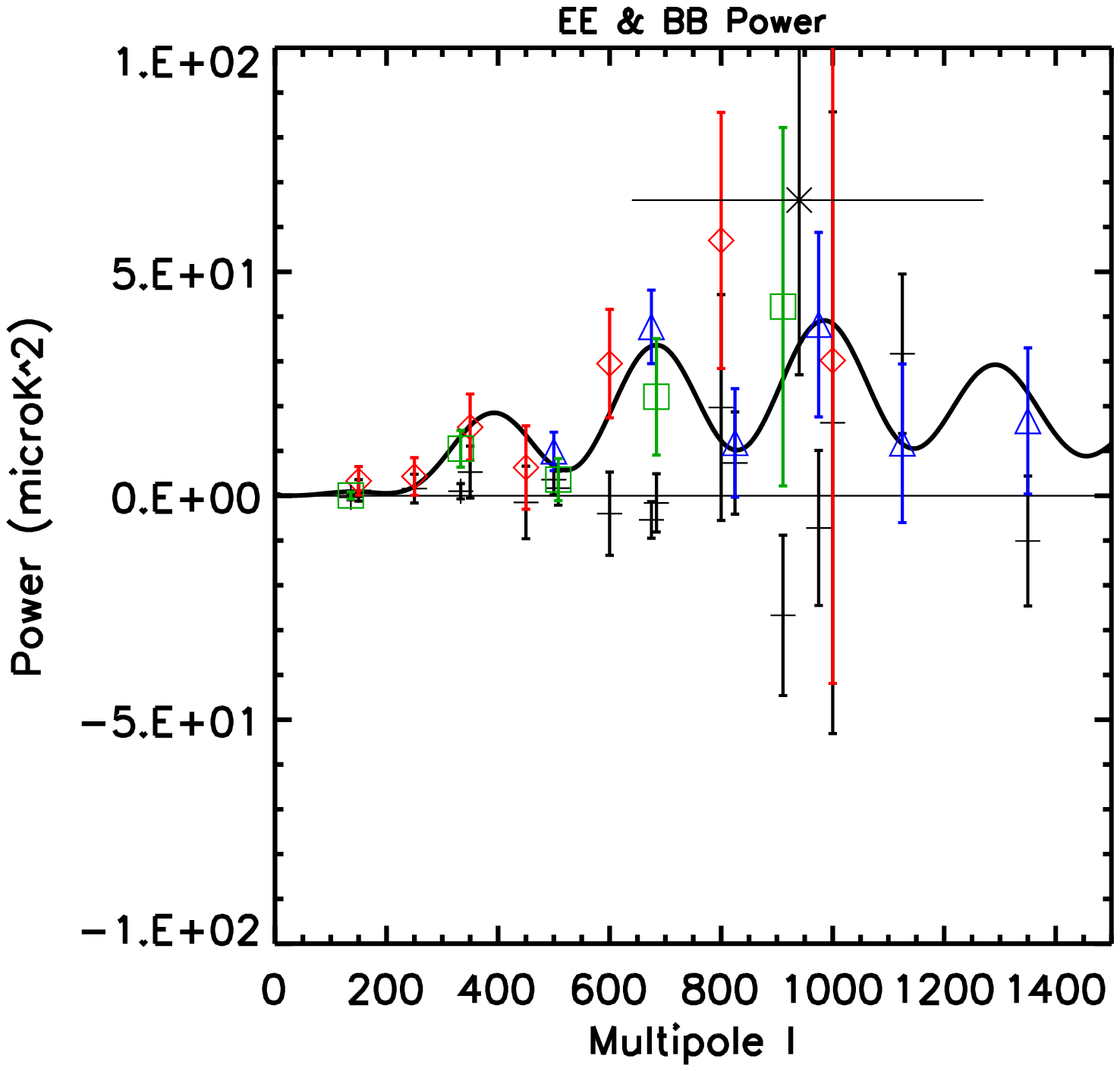}
\caption{\label{fig:ee}EE and BB power spectra.  The curve 
shows the standard model prediction for $\CEE$.  Points are labeled 
as for the previous figure; the black asterisk is the
CAPMAP $E$--mode power measurement.  
Here, the thin black error bars give
each experiment's $B$--mode measurements (all consistent with zero).}
\end{minipage} 
\end{figure}

\section{Observational Effort} 
We often refer to polarization as the {\em next step} in CMB science. 
While appropriate, this erroneously gives the impression that it remains
for the future, when in fact, different experiments have already measured
CMB polarization. I give a summary in Figures~\ref{fig:te} and \ref{fig:ee}.

The DASI experiment at the South Pole was the first to detect CMB
polarization, both $E$ and $TE$ modes~\cite{dasi_det}; their recently 
published 3--year results~\cite{dasi_3y} are shown in 
Figures~\ref{fig:te} and \ref{fig:ee} as the green squares.   
The original DASI detection was followed by WMAP's
measurement of $\CTE$ on large scales down to 
$l\sim 500$ from the first year data~\cite{wmap_polar}; these 
are not reproduced here. 

More recently, the BOOMERanG collaboration reports measurements of $\CTT$,
$\CTE$ and $\CEE$ and a non--detection of
$B$--modes~\cite{boom_polar}. Combining their new
BOOMERanG data with other CMB and large scale structure data,
MacTavish et al.~\cite{boom_params} constrain $r<0.36$ (95\%).  
These results are shown in Figures~\ref{fig:te} and \ref{fig:ee} 
as the red diamonds. 

The CBI experiment has also published new measurements of $\CTT$,
$\CTE$ and $\CEE$, as well as a non--detection of $B$--
modes~\cite{cbi_polar}.  These
are shown in Figures~\ref{fig:te} and \ref{fig:ee} as the blue triangles.  
Finally, the black asterisk in Figure~\ref{fig:ee} gives the 
$E$--mode power measurement by CAPMAP~\cite{capmap}. All of these 
results are consistent with each other and with the prediction of the
standard cosmological model assuming pure adiabatic modes, shown
in the figures as the black curve.

Although these new $B$--mode limits are still far from placing any 
important constraints on either gravity waves or lensing, the results are
significant for what they imply about Galactic foregrounds.  We expect
these foregrounds to generate $E$-- and $B$--modes with equal strength -- 
there is no symmetry preferring one over the other.  The lack of $B$--mode
power thus suggests that foreground contamination in these
data sets is well below the measured $E$ and $TE$ signals.

Scheduled for launch in 2007/2008, the Planck satellite will 
greatly advance our knowledge of CMB polarization by providing
foreground/cosmic variance--limited measurements of $\CTE$ and 
$\CEE$ out beyond $l\sim 1000$.  We also expect to detect the
lensing signal, although with relatively low precision, and could
see gravity waves at a level of $r\sim 0.1$.  The Planck blue book 
quantifies these expectations.  

A leap in instrument sensitivity is required in order to go beyond 
Planck and get at the $B$--modes from lensing and gravity waves.
This important science is motivating a vast effort world wide at developing a 
new generation of instruments based on large detector arrays.  Numerous
ground--based and ballon--borne experiments are actually observing 
or being prepared.  In the longer term future, both NASA (Beyond Einstein)
and ESA (Cosmic Vision) have listed a dedicated CMB polarization mission
as a priority in the time frame 2015-2020.  Such a mission could reach
the cosmic variance limit on the lensing power spectrum to measure 
the neutrino mass scale and perhaps detect primordial gravity waves from 
inflation near the GUT scale.  The exciting journey has begun.

\end{document}